# Further Evaluations of a Didactic CPU Visual Simulator (CPUVSIM)

**Renato Cortinovis**
Freelance Researcher
Italy
rmcortinovis@gmail.com

**Tamer Mohamed Abdellatif**
Canadian University Dubai, United Arab Emirates
tamer.mohamed@cud.ac.ae

**Devender Goyal**
Raytheon Technologies, USA
dg1998@gmail.com

**Luiz Fernando Capretz**
Western University
Canada
lcapretz@uwo.ca

**Abstract**
This paper discusses further evaluations of the educational effectiveness of an existing CPU visual simulator (CPUVSIM). The CPUVSIM, as an Open Educational Resource, has been iteratively improved over a number of years following an Open Pedagogy approach, and was designed to enhance novices' understanding of computer operation and mapping from high-level code to assembly language. The literature reports previous evaluations of the simulator, at K12 and undergraduate level, conducted from the perspectives of both developers and students, albeit with a limited sample size and primarily through qualitative methods. This paper describes additional evaluation activities designed to provide a more comprehensive assessment, across diverse educational settings: an action research pilot study recently carried out in Singapore and the planning of a more quantitative-oriented study in Dubai, with a larger sample size. Results from the pilot study in Singapore confirm the effectiveness and high level of appreciation of the tool, alongside a few identified challenges, which inform the planning of the more comprehensive evaluation in Dubai.

## 1. Introduction

Numerous CPU visual simulators have emerged over time with the primary objective of enhancing the understanding of computer operation (Nikolic et al., 2009). These simulators cater to various levels of expertise and often specialize in specific facets of computer science, such as computer security (Imai et al., 2013) or pipelining (Zhang and Adams III, 1997). Among this diverse array of simulators, a small subset addresses a well-recognized issue: that students – despite studying both high-level programming languages and computer architecture fundamentals – frequently struggle to grasp how high-level code actually executes on computer hardware (Evangelidis et al., 2021; Miura et al., 2003). These concepts are considered fundamental in computer science and software engineering education and training: the Software Engineering Body of Knowledge (Bourque and Fairley, 2022), for example, reports that "software engineers are expected to know how high-level programming languages are translated into machine languages".

In this context, the CPUVSIM (Cortinovis, 2021) – an Open Educational Resource available from Merlot and OER Commons – supports novices in comprehending the fundamental components of a simplified CPU, and in understanding the mapping from high-level control structures to low-level code, i.e. assembly and machine code. This is achieved through detailed animations that illustrate the execution of instructions, and empowering learners to write meaningful programs using a minimalist yet representative assembly language.

Figure 1 shows a screenshot of the CPUVSIM running in a browser, with a simple program loaded in RAM. The user can execute the program one instruction or micro-instruction at a time, at the desired speed. The user can interactively modify, at any time, the content of the RAM, or any register in the CPU. While the execution is animated, a voice over explains what is happening – in English, Spanish, or Italian.

As detailed by Cortinovis (2021), the development of CPUVSIM sought to address limitations observed in existing applications. While some simulators, such as LMC (Higginson, 2014), were deemed overly simplistic, others were considered unnecessarily complex. CPUVSIM, on the other hand, was honed through iterative improvements and extensions, building upon the foundation of an already popular visual simulator known as PIPPIN (Decker and Hirshfield, 1998). Its development process followed a sustainable Open Pedagogy approach (Wiley and Hilton, 2018) in the form of non-disposable assignments to computer science students over multiple years.

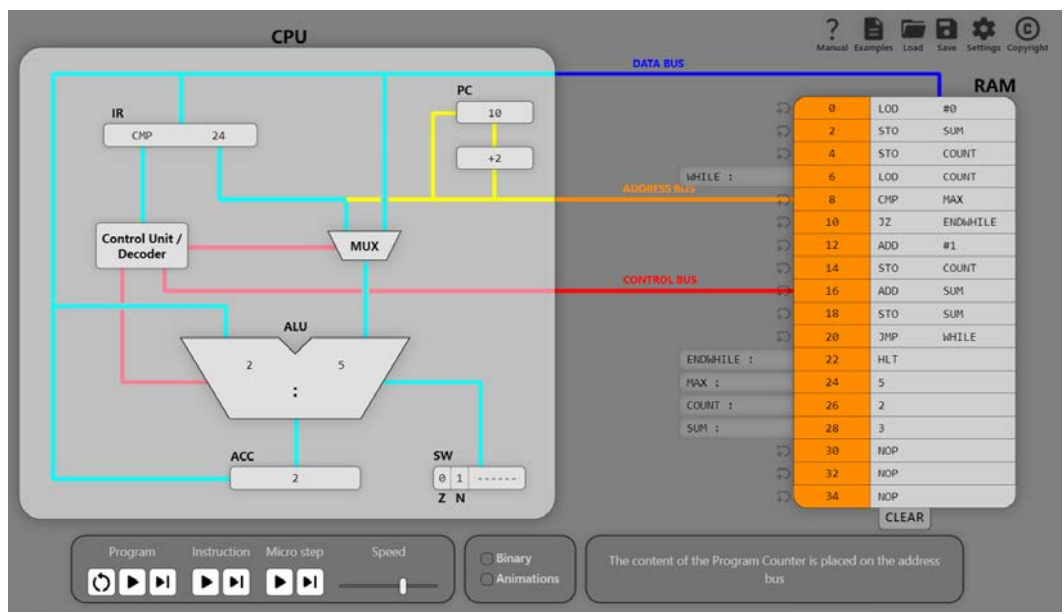

*Figure 1 - A screenshot of the CPUVSIM*

The CPUVSIM has previously undergone evaluations of its educational effectiveness, albeit with limited sample size and primarily through qualitative methods. These evaluations were conducted from the perspectives of both its developers and students who engaged with it in different contexts. In this paper, we describe an evaluation recently conducted in Singapore and also describe an evaluation planned to be conducted in Dubai. These studies outline additional evaluation activities designed to provide a more comprehensive assessment of the CPUVSIM. Our efforts encompass diverse settings, employing as far as possible complementary approaches.

Following the general introduction in this section, Section 2 reports on the evaluation of CPU simulators in the literature, Section 3 describes our Action Research pilot evaluation recently carried out in Singapore and its findings. Section 4 describes the planning of the evaluation to be conducted at a University in Dubai with a larger number of students, utilizing again Action Research but complemented with elements of a quantitative-oriented experimental approach. Finally, Section 5 presents our conclusions.

## 2. CPU simulators evaluation strategies

As mentioned earlier, the CPUVSIM has already undergone some limited and mainly qualitative evaluations. Cortinovis (2021) describes its informal qualitative evaluation from the developers' point of view, who deeply appreciated, in particular, the opportunity to work on a real problem in a real context, and the opportunity to contribute to the common good. Cortinovis and Rajan (2022) describe the evaluation from the students' point of view, both in two specialized technical schools in Italy (K12 and lifelong adult education) and in a first and a second-year undergraduate computer architecture courses in Colorado (USA). The students who used the simulator provided very positive feedback, which was analysed with a qualitative thematic content analysis, and was then used to further improve and extend the latest version of the simulator.

Nikolic et al. (2009), evaluate a rich set of existing CPU simulators, but only on the basis of characteristics identified from the documentation. Some of the criteria they used, such as level of coverage of the topics included in standard curricula, are not considered fully relevant in this context: the CPUVSIM is meant to support a firm grasp of the fundamental mechanisms, but at a relatively high level of abstraction, without dwelling too much in details.

Imai et al. (2013) evaluated the correlation between tests carried out on their simulator and the course final exam results. The strong correlation they reported is interesting, but to demonstrate the effectiveness of their simulator, it is necessary to compare it with an alternative tool or a comparable teaching method. Indeed, in subsequent works (Imai et al., 2018), they adopted a qualitative approach alone with a simple questionnaire.

Chalk (2002) and Mustafa (2010) both used a mixed qualitative and quantitative strategy. The quantitative approach, in particular, makes use of a quasi-experimental schema, with experimental and control groups, and pre and post knowledge tests. Although the experimental strategy is instrumental in collecting supporting evidence about the effectiveness of the tool, and while pre- and post- knowledge tests can demonstrate the improvements of students' knowledge using the simulator, this approach does not provide information about its effectiveness against alternative strategies. Chalk (2002) demonstrates, in particular, the importance of referring to precisely formulated learning objectives, to test results against.

## 3. CPUVSIM pilot evaluation in Singapore

We planned and executed a first pilot evaluation of the simulator, on a small scale (13 students in total), in two undergraduate courses at the Yale-NUS College in Singapore: a course on C Programming and a course on Software Verification and Validation. The simulator was used in the first course to help students understand the mapping between C control structures and assembly code. It was used in the second course to test programs at machine language level. Considering the limited number of students, and the limited possibilities to control the many factors involved (different classes, different teachers, etc.), we considered it appropriate to adopt a socially-oriented, situational Action Research methodology, preferring a more postmodernist-oriented approach over a strictly positivist one (Kemmis and McTaggart, 2000).

Given the overall goal of grasping how code written in high-level language is actually executed on the hardware of a computer, we outlined first, as recommended by Chalk (2002), the learning objectives:

- Understand the role of the key components of a CPU.
- Understand the mapping from high-level to low-level control structures (assembly and machine) code.
- Code meaningful high-level programs with a minimalist but representative assembly language.

More specifically:

- Describe typical assembly instructions supported by a CPU.
- Explain the fundamental steps carried out by the main subcomponents of a CPU, to execute a given assembly instruction.
- Identify the information transferred on the Data bus, Address bus, and Control bus during each step of every instruction.
- Apply the suitable numeric/immediate and direct addressing modalities.
- Exemplify the use of the CPU flags through simple examples.
- Translate a program in C with a single control structure to assembly code.

According to the adopted research methodology, we defined an action plan for the proposed intervention, including specific pedagogical activities as well as "Data analysis and critical reflection", and "Refinement of the planned intervention for future courses".

In particular, we foresaw a first activity to present in class the CPUVSIM and its associated e-book (1.5 hours), follow-up students' activities to be started in class and completed at home (a couple of hours to familiarize individually with the CPUVSIM and related educational material), plus an additional hour to complete the graded activities. These included:

> Briefly explain the differences between conditional and non-conditional jump.

> Briefly list/describe the steps carried out by the main sub-components of a CPU, to execute the instructions ADD #20 and ADD 20 (immediate and direct addressing).

Identify the missing instruction in the following translation of an IF-THEN-ELSE control structure to assembly:

| IF SUM == 2<br>  THEN SUM=3<br>  ELSE SUM=5<br>ENDIF |        LOD SUM<br>       CMP #2<br>       JNZ ELSE<br>       LOD #3<br>         // MISSING CODE?<br>ELSE:   LOD #5<br>ENDIF: STO SUM<br>       HALT<br>SUM:   0 |
|---|---|

The second course on software verification and validation included the following additional assignments:

> Use the Simulator to test if the translation of the following high-level control structures to Assembler are correct or not […]; explain your answer.
>
> Discuss Specific Testing Strategies for Assembly code.

We finally specified a survey with Likert type questions and open questions (Mustafa, 2010; Cortinovis and Rajan, 2022), to collect feedback about the CPUVSIM and the learning experience, such as:

> What ameliorations could be made to the simulator and/or related e-book to improve your learning experience?

The final assignment was graded and analysed with psychometric Classical Test Theory (Novick, 1996). Taking into account the limited number of students, the Likert-type questions in the survey were analysed with basic descriptive statistics (Mustafa, 2010), the open questions in the survey were analysed with thematic content analysis (Cortinovis and Rajan, 2022).

Finally, we carried out a critical reflection on the effectiveness of the intervention to derive the recommendations for planning the subsequent intervention – according to the iterative nature of Action Research, and to the goal of a pilot.

### 3.1 Pilot Results

The data extracted from the survey (Table 1) on 13 students shows that the CPUVSIM was definitely appreciated, especially for understanding how C control structures actually get executed on a computer, which was the main goal.

| Questions | Strongly Agree + Agree (%) | Strongly Disagree + Disagree (%) | Neutral (%) |
|---|---|---|---|
| The simulator and related e-book were motivating and interesting. | 77% (10) | 0% | 23% (3) |
| The simulator and related e-book were useful for understanding how C control structures actually get executed on a computer. | 85% (11) | 0% | 15% (2) |
| I found the simulator too complicated to understand and use effectively. | 23% (3) | 46% (6) | 31% (4) |

*Table 1 – Sample extracted from the survey in Singapore.*

Interestingly, a relevant number (23%) of students stated that the simulator was not easy to understand and use effectively: this was probably due to the limited time devoted to its presentation (just 1.5 hours in total). Indeed, a first student who found the simulator too complicated suggested having "more hand holding in class"; a second student considered that "the lecture included too much". A student found that the simulator was too fast: "It was challenging keeping up with its fast-pace while still understanding newly introduced concepts". Obviously, this student did not notice the possibility to control the speed, which was explicitly appreciated by other students.

Despite these problems, the students' answers on the assignments demonstrated a remarkable grasp of the targeted key concepts. There were no incorrect solutions to the assignments, even if there were omissions of relevant details in a few of them – notably from a student who found the use of the simulator somewhat complicated.

These overall positive outcomes were strongly correlated with the students' self-perceptions: one of them, for example, stated that she reached "a solid understanding of how high-level code runs on hardware", while another one found it "really eye-opening to see how it actually works at the base level". The number of students involved in this pilot was limited, yet the evaluation results confirm previous results available in the literature (Cortinovis and Rajan, 2022). The main lesson learned for future deliveries of the course, is the need to dedicate more time to coaching the students in the use of the simulator, so that all of them can get the most from it.

**4. CPUVSIM planned evaluation in Dubai**
In Dubai we aim to improve the previous evaluations addressing two potentially weak aspects of action research: generalizability and rigor. Concerning generalizability, we are evaluating the simulator in different contexts, that is, different courses, multiple classes, and different countries. To improve the rigor of the evaluation process, we take advantage of the larger sample size available (120+ students), enriching the qualitative-oriented action research design used in Singapore, with a quantitative-oriented experimental approach.

Drawing lessons from the pilot evaluation conducted in Singapore, we will allocate additional time to ensure that every student gains complete mastery over the utilization of CPUVSIM and its accompanying documentation. First, we will dedicate a decent time to our lab instructors to train on and master the simulator. This is planned to take place during the pre-semester preparation period of two weeks. During this period, the course instructors, with the support of the lab instructors, will work on integrating the simulator within the course syllabi and preparing the simulator-based assessment tasks. Accordingly, full two lab sessions (2 hours each) will be dedicated to the students' training on the simulator. The first lab session (2 hours) will be dedicated to introducing the simulator's built-in CPU instructions in addition to the education supporting features, such as the instructions execution simulator and the simulator e-book. After finishing this lab session, the students will be left with a simulator-based homework. In the second lab session (2 hours), the students will be provided with time dedicated to homework discussion and one-by-one coaching on the answer of each of the homework tasks using the simulator. This way, the students will acquire a fair understanding level of the simulator's functionality before proceeding with more challenging tasks.

Furthermore, we will include the following additional learning objective:

- Translate a program in a high-level language with multiple control structures, both sequenced and nested, to assembly code.

Therefore, beyond the exercises proposed in Singapore, we will conduct a more comprehensive assessment of students' proficiency in translating high-level constructs, such as loops, logic operators, and arithmetic operations, into assembly language. For example:

- write an assembly program that determines whether the value stored in a variable "var1" is odd or even;
- write an assembly program that performs a comparison between two signed variables. Var1=7Fh and Var2=80h. Then saves the highest and the lowest variables in HIGH and LOW variables respectively;
- write an assembly program that determines whether a given positive integer number satisfies the Collatz conjecture.

4.1 Experimental Design and Population
We have a total of 120+ students that will be partitioned into two groups. The first one will follow the traditional educational path of the previous years, which did not use a simulator but was based on theoretical lessons and paper-based exercises. The second one will follow an educational path modified with the new intervention, using the CPUVSIM. The students, who have Python and Java programming background, will have different lab instructors and teaching assistants.

### 4.2 Quantitative Data Collection and Analysis
We will gather quantitative data through the following assessment setups:

- Exercises under an invigilated environment: for this setup, students will be asked to answer the assessments at the lab using our learning management system (LMS). The students will have no access to the internet. The assessment will have a specified time limit and the LMS system will record the time taken by the student to complete each task (completing the task in less time will lead to collecting more marks). We plan to conduct this assessment setup twice within the course timeline.
- Group-based exercises carried out in the lab: this kind of assessment will allow the students in small sub-groups (3-4 students) to hone their assembly level programming and benefit from each other. In case of any needed support from the instructor, the students will be asked to submit their inquiries via the LMS system.
- Written exams: this includes both graded mid-term and final exams. Both students' grades and specific mistakes will be gathered for this kind of assessment setup.

Therefore, in addition to the students' grades, we will gather other quantitative data such as the time needed to finish the assessment, number of students' mistakes, types/categories of the students' mistakes (for instance, whether due to incorrect understanding of the logic of jump instructions or to incorrect understanding of the mapping between high level programming and assembly code logic), and the Grade Point Average of each group member. Furthermore, each assessment's activity/question will be mapped to a certain learning objective. This allows us to evaluate the effectiveness of using the simulator on improving the students' knowledge for each of our learning objectives, in addition to the overall impact on the students' performance along course(s).

We will adopt ANOVA variance analysis statistical test to verify whether the results for the two main student groups show an overall statistically significant difference based on the use of CPUVSIM according to the sample size at hand. In addition, we plan to apply the Tukey's Honestly Significant Difference (HSD) statistical test to figure out which group of data parameters is impacted the most by using the CPUVSIM based on the sample at hand. Examples of data parameters that can be studied by HSD are: time needed to finish the task, number of mistakes.

### 4.3 Qualitative Data Collection and Analysis
Qualitative data will be collected using a survey with both Likert-type questions and open questions similar to the ones adopted in the pilot study. Here, however, we plan to extend our qualitative data by collecting feedback from teachers too. Therefore, the survey questions for the teachers will include, for example: How does the CPUVSIM impact on the students' understanding of the assembly language structure? How does it impact related explanations? Would you recommend making use of the CPUVSIM, and how?

We plan to interview all the course instructors as well as one student from each student-focused sub-group. To analyse these qualitative data, both thematic and narrative analysis will be adopted. Thematic analysis will help in identifying and interpreting the patterns from our survey results, while narrative analysis will provide a better understanding of the motivation behind the feedback provided by the interviewees.

Finally, we will also run a longitudinal study, because we suspect that the students with hands-on experience with the simulator might better retain over time the competences acquired, compared to the students who followed the more traditional path. Therefore, we will retest the students of both groups after 12 months, to assess the possible different levels of retention of key concepts and competencies.

### 4.4 Threats to Validity
Generalizability and rigor are the two main weak aspects of the situational nature of action research. Yet, the variety of contexts where these evaluations are carried out (Singapore) and planned (Dubai), in addition to the evaluations previously reported in the literature (Italy and USA), should contribute to support the generalization of the outcomes. Additionally, we integrated in the methodology for the planned evaluation in Dubai, a quantitative-oriented experimental component to improve rigor: the use of complementary research methodologies, selected to better fit the particular contexts, should help compensate for their weaknesses.

More importantly, we acknowledge our limited control over numerous variables, particularly the inevitably diverse approaches employed by various educators when utilizing the tool. The challenge lies in discerning the tool's impact amidst the multitude of factors influencing student learning. It is imperative to recognize the tool as just a component within a broader socio-technical system, as highlighted by Mulholland (personal communication, 2023). To tackle this challenge, our evaluation strategy in Dubai includes qualitative assessments from educators' perspectives. These assessments aim to glean insights into how teachers perceive the CPUVSIM and its effects on their educational endeavours. Additionally, we leverage the recently developed CPUVSIM accompanying e-book titled "A Gentle Introduction to the Central Processing Unit (CPU) and Assembly Language". This Open Educational Resource, available via Merlot or OER Commons, offers some pedagogical support for utilizing the CPUVSIM. It provides explanations and practical activities to support both teachers as well as self-learners. Notably, this interactive e-book integrates the CPUVSIM seamlessly. Each programming example or exercise within the e-book features live images, embedding the fully functional CPUVSIM. Users can execute, modify, and re-execute these "images" at will, enhancing the interactive learning experience.

### 4.5 Ethical Considerations
In our plan, only one of the two groups of students will have the opportunity to reap the potential advantages of using the CPUVSIM, which can be questioned from the ethical point of view. To overcome this problem, in case the evaluation would show that a group was considerably disadvantaged compared to the other, we plan to offer, at the end of the evaluation, some extra educational activities to level their competences. These extra activities would target specifically the topics where the evaluation might have identified significant differences.

## 5. Conclusions
We have outlined our plans for assessing the educational effectiveness of the CPUVSIM simulator. The feedback from the pilot in Singapore, using Action Research, confirms a positive effect on students' ability to grasp important concepts and good appreciation for the tool, as reported in the literature, together with the identification of some challenges. This provided useful indications for the wider evaluation planned in Dubai, where we will enrich the qualitative Action Research methodology with a more quantitative-oriented study, aiming to address concerns about generalizability and rigor.

Through these activities, we are collecting feedback from students and teachers in diverse geographical regions, broadening the perspective on how the CPUVSIM resonates with stakeholders from different cultural backgrounds and educational systems.

## 6. References

Bourque, P., & Fairley, R. E. (2022). Guide to the Software Engineering Body of Knowledge, Version 4.0 beta. IEEE Computer Society.

Chalk, B. (2002). Evaluation of a Simulator to Support the Teaching of Computer Architecture. In 3rd Annual LTSN-ICS Conference, Loughborough University.

Cortinovis, R. (2021). An educational CPU visual simulator. 32nd Annual Workshop of the Psychology of Programming Interest Group.

Cortinovis, R., & Rajan, R. (2022). Evaluating and improving the educational CPU visual simulator: a sustainable open pedagogy approach. In Proceedings of the 33rd Annual Workshop of the Psychology of Programming Interest Group, 189-196.

Decker, R., & Hirshfield, S. (1998). The Analytical Engine: An Introduction to Computer Science Using the Internet. PWS Publishing, Boston.

Evangelidis, G., Dagdilelis, V., Satratzemi, M., & Efopoulos, V. (2021). X-compiler: yet another integrated novice programming environment. In Proceedings of the IEEE International Conference on Advanced Learning Technologies, 166-169.

Higginson, P. (2014). Little Man Computer [Javascript application]. Retrieved September 2023, from https://peterhigginson.co.uk/LMC/.

Imai, Y., Hara, S., Doi, S., Kagawa, K., Ando, K., & Hattori, T. (2018). Application and evaluation of visual CPU simulator to support information security education. IEEJ Transactions on Electronics, Information and Systems. 138, 9, 1116-1122.

Imai, Y., Imai, M., & Moritoh, Y. (2013). Evaluation of visual computer simulator for computer architecture education. International Association for Development of the Information Society.

Kemmis, S., & McTaggart, R. (2000). Participatory action research. In Handbook of Qualitative Research, edited by Norman K. Denzin & Yvonna S. Lincoln, 2nd ed., 567-605.

Miura, Y., Keiichi, K., & Masaki, N. (2003). Development of an educational computer system simulator equipped with a compilation browser. In Proceedings of the International Conference of Computers in Education, 140-143.

Mustafa, B. (2010). Evaluating a system simulator for computer architecture teaching and learning support. Innovation in Teaching and Learning in Information and Computer Sciences. 9, 1, 100-104.

Nikolic, B., Radivojevic, Z., Djordjevic, J., & Milutinovic, V. (2009). A survey and evaluation of simulators suitable for teaching courses in computer architecture and organization. IEEE Transactions on Education. 52, 4, 449-458.

Novick, M. R. (1996). The axioms and principal results of classical test theory. Journal of Mathematical Psychology. 3, 1, 1-18.

Wiley, D., & Hilton, J. (2018). Defining OER-enabled pedagogy. International Review of Research in Open and Distance Learning. 19, 4.

Zhang, Y., & Adams III, G. B. (1997). An interactive, visual simulator for the DLX pipeline. IEEE Computer Society Technical Committee on Computer Architecture Newsletter. 9-12.